\documentstyle[12pt]{article}
\begin{document}
\pagestyle{empty}
\setlength{\oddsidemargin}{0.5cm}
\setlength{\evensidemargin}{0.5cm}
\setlength{\footskip}{1.5cm}
\renewcommand{\thepage}{-- \arabic{page} --}
% -------------------------------------------------------------------
\vspace*{-2.5cm}
\begin{flushright}
TOKUSHIMA 95-01 \\ (hep-ph/9501355) \\ January 1995
\end{flushright}
\vspace*{1.25cm}

\renewcommand{\thefootnote}{*)}
\centerline{\ {\large\bf Non-trivial Tests of EW Corrections via
$\!\!\mbox{\boldmath $\alpha$}$,
$\mbox{\boldmath $G$}_{\!\mbox{\footnotesize\boldmath $F$}}$ and
$\mbox{\boldmath $M$}_{\!\mbox{\footnotesize\boldmath $W$}\!\!,\,
\mbox{\footnotesize\boldmath $Z$}}$}\ \footnote{Talk presented at INS
Workshop ``{\it Physics of $e^+e^-$, $e^-\gamma$ and $\gamma\gamma$
collisions at linear accelerators}", Inst. for Nucl. Study (INS),
Univ. of Tokyo, Japan, December 20-22, 1994 (to appear in the
Proceedings).}}

\vspace*{1.75cm}

\renewcommand{\thefootnote}{**)}
\centerline{\sc \phantom{**)}Zenr\=o HIOKI\,\footnote{E-mail:
hioki@ias.tokushima-u.ac.jp}}

\vspace*{1.75cm}

\centerline{\sl Institute of Theoretical Physics,\ University of
Tokushima}

\vskip 0.3cm
\centerline{\sl Tokushima 770,\ JAPAN}

\vspace*{3cm}

\centerline{ABSTRACT}

\vspace*{0.4cm}
\baselineskip=20pt plus 0.1pt minus 0.1pt
The standard electroweak theory is tested at non-trivial quantum
correction level through $\alpha$, $G_F$ and the latest data of the
weak-boson masses. The improved-Born approximation and the
non-decoupling top-quark effects are studied without depending on the
CDF data of $m_t$, while the bosonic effects are examined by fully
taking account of it.

\vfill
% -------------------------------------------------------------------
\newpage
\pagestyle{plain}
\renewcommand{\thefootnote}{\sharp\arabic{footnote}}
\setcounter{footnote}{0}
\baselineskip=21.0pt plus 0.2pt minus 0.1pt

Many particle physicists now believe that the standard electroweak
theory (plus QCD) describes correctly phenomena below $O(10^2)$ GeV.
In fact, there has been observed no discrepancy between experimental
data and the corresponding predictions by this theory with radiative
corrections. Novikov et al. claimed in \cite{NOV}, however, that the
Born approximation based on $\alpha(M_Z)$ instead of $\alpha$
(``improved-Born" approximation) explains all electroweak precision
data up to 1993 within the $1\sigma$ accuracy, where $\alpha$ and
$\alpha(M_Z)$ are the QED coupling constants at $m_e$ and $M_Z$
scales respectively. This means that the electroweak theory had not
been tested by that time at ``non-trivial" level.

After their work, a new experimental value of $M_W$ was reported
($M_W^{exp}=80.23\pm 0.18$ GeV) \cite{DFHKS}, and furthermore CDF
collaboration at FNAL tevatron collider obtained some evidence on the
top quark ($m_t^{exp}=174\pm 17$ GeV) \cite{TOP}. Being stimulated
by them, I started to study the present issue, and worked up the
results into three papers \cite{ZH941,ZH942,ZH943}. At this workshop,
I showed the main point of these works.

What I studied is ``structure of EW(electroweak) corrections''. The
EW corrections consist of several parts with different properties,
and I examined via $\alpha$, $G_F$ and $M_{W,Z}$ what would happen if
each of them would not exist. More concretely, I examined whether the
improved-Born approximation still works or not, and then focused on
the top-quark contribution which does not decouple, i.e., becomes
larger and larger as $m_t$ increases. It is very significant to test
it because the existence of such effects is a characteristic feature
of theories in which particle masses are produced through spontaneous
symmetry breakdown plus large Yukawa couplings. Furthermore, I also
studied the bosonic contribution to the whole corrections. From a
theoretical point of view, this is another important test since the
bosonic part includes the gauge-boson- and Higgs-boson-loop effects.

Through the $O(\alpha)$ corrections to the muon-decay amplitude,
$\alpha$, $G_F$ and $M_{W,Z}$ are connected as
\begin{eqnarray}% ---------------------------------------------------
M_W^2={1\over 2}M_Z^2
\biggl\{ 1+
\sqrt{\smash{1-{{2\sqrt{2}\pi\alpha}
\over{M_Z^2 G_F (1-{\mit\Delta}r)}}}
\vphantom{A^2\over A}
}~\biggr\}. \label{eq1}
\end{eqnarray}% -----------------------------------------------------
Here ${\mit\Delta}r$ expresses the corrections, and it is a function
of $\alpha$, $G_F$, $M_Z$, $m_f$ and $m_{\phi}$. This formula,
the $M_W$-$M_Z$ relation, is the main tool of my analyses.\footnote{
    Strictly speaking, Eq.(\ref{eq1}) is not complete: It is a
	formula based on the one-loop calculations (with resummation of
	the leading-log terms by the replacement $(1+{\mit\Delta}r)$
	$\rightarrow$ $1/(1-{\mit\Delta}r)$). Over the past several
	years, some corrections beyond the one-loop approximation have
	been computed. They are two-loop top-quark corrections
	\cite{BBCCV} and QCD corrections up to $O(\alpha_{\rm QCD}^2)$
	\cite{HKl} for the top-quark loops. As a result, we have now a
	formula including $O(\alpha\alpha_{\rm QCD}^2 m_t^2)$ and
	$O(\alpha^2 m_t^4)$ effects. In	the following, $M_W$ is always
	computed by incorporating all of these higher-order terms as
	well, although I will express the whole corrections with these
	terms also as ${\mit\Delta}r$ for simplicity.}\ %----------------
Before proceeding to the actual analyses, let me show by using this
formula how the theory with the full corrections is successful,
although it is already a well-known fact. The $W$-mass is computed
thereby as
\begin{eqnarray}% ---------------------------------------------------
M_W^{(0)}=80.941\pm 0.005\ {\rm GeV\ \ and}\ \
M_W=80.33\pm 0.11\ {\rm GeV}
\end{eqnarray}% -----------------------------------------------------
for $M_Z^{exp}=91.1888\pm 0.0044$ GeV \cite{Sch}, where $M_W^{(0)}$
and $M_W$ are those without and with the corrections respectively,
and $M_W$ is for $m_t^{exp}=174 \pm 17$ GeV \cite{TOP},
$m_{\phi}=300$ GeV and $\alpha_{\rm QCD}(M_Z)$=0.118. We can find
that the theory with the corrections is in good agreement with the
experimental value $M_W^{exp}=80.23\pm 0.18$ GeV, while the tree
prediction fails to describe it at more than 3.9$\sigma$ (99.99
\% C.L.).

We are now ready. First, it is easy to see if taking only
$\alpha(M_Z)$ into account is still a good approximation. The
$W$-mass is calculated within this approximation by putting ${\mit
\Delta}r=0$ and replacing $\alpha$ with $\alpha(M_Z)$ in
Eq.(\ref{eq1}), where $\alpha(M_Z)=1/(128.87\pm 0.12)$ \cite{Jeg}.
The result is
\begin{eqnarray}% ---------------------------------------------------
M_W[{\rm Born}] =79.957\pm 0.017~{\rm GeV},
\end{eqnarray}% -----------------------------------------------------
which leads to
\begin{eqnarray}% ---------------------------------------------------
M_W^{exp}-M_W[{\rm Born}]~=~0.27\pm 0.18~{\rm GeV}.
\end{eqnarray}% -----------------------------------------------------
This means that $M_W[{\rm Born}]$ is in disagreement with the data
now at $1.5\sigma$, which corresponds to about 86.6 \%\ C.L..
Although the precision is not yet sufficiently high, it indicates
some non-Born terms are needed which give a positive contribution to
the $W$-mass. It is noteworthy since the electroweak theory predicts
such positive non-Born type corrections unless the Higgs is extremely
heavy (beyond TeV scale). A similar result was obtained also in
\cite{NOV94}.

The next test is on the non-decoupling top-quark effects. Except for
the coefficients, their contribution to ${\mit\Delta}r$ is
\begin{eqnarray}% ---------------------------------------------------
{\mit\Delta}r[m_t]\sim \alpha (m_t/M_Z)^2+\alpha\ln(m_t/M_Z).
\end{eqnarray}% -----------------------------------------------------
According to my strategy, I computed the $W$-mass by using the
following ${\mit\Delta}r'$ instead of ${\mit\Delta}r$ in
Eq.(\ref{eq1}):
\begin{eqnarray}% ---------------------------------------------------
{\mit\Delta}r'\equiv {\mit\Delta}r-{\mit\Delta}r[m_t].
\end{eqnarray}% -----------------------------------------------------
The resultant $W$-mass is denoted as $M_W'$. The important point is
to subtract not only $m_t^2$ term but also $\ln(m_t/M_Z)$ term,
though the latter produces only very small effects unless $m_t$ is
extremely large. ${\mit\Delta}r'$ still includes $m_t$ dependent
terms, but no longer diverges for $m_t\to +\infty$ thanks to this
subtraction. I found that $M_W'$ takes the maximum for the largest
$m_t$ and the smallest $m_{\phi}$. That is, we get an inequality
\begin{eqnarray}% ---------------------------------------------------
M_W'\ \leq\ M_W'[m_t^{max}, m_{\phi}^{min}],
\end{eqnarray}% -----------------------------------------------------
which holds for any experimentally-allowed values of $m_t$ and
$m_{\phi}$.

Although the CDF report on the top-quark is quite exciting, but its
final establishment must come after D0 collaboration confirms it.
Therefore, I took a conservative position and calculated the
right-hand side of the above inequality for $m_t^{max}\to +\infty$.
Concerning $m_{\phi}^{min}$, on the other hand, we can use the
present experimental bound $m_{\phi}^{exp}>61.5$ GeV \cite{Kob}. The
accompanying uncertainty for $M_W'$ is estimated at most to be about
0.03 GeV. We have then
\begin{eqnarray}% ---------------------------------------------------
M_W' < 79.865 (\pm 0.030) \ {\rm GeV\ \ and\ \ }
M_W^{exp}-M_W' > 0.36\pm 0.18\ {\rm GeV},
\end{eqnarray}% -----------------------------------------------------
which show that $M_W'$ is in disagreement with $M_W^{exp}$ at more
than $2.0\sigma$ (=95.5 \% C.L.). This means that 1) the electroweak
theory is not able to be consistent with $M_W^{exp}$ \underline{
whatever values $m_t$ and $m_{\phi}$ take} if the non-decoupling
top-quark corrections ${\mit\Delta}r[m_t]$ would not exist, and 2)
the theory with ${\mit\Delta}r[m_t]$ works well, as shown before, for
experimentally-allowed $m_t$ and $m_{\phi}$.\footnote{Of course, it
    is conservative in this case to {\it use} the CDF data.}\ %------
Combining them, we are led to an interesting phenomenological
indication that the latest experimental data of $M_{W,Z}$ demand,
\underline{independent of $m_{\phi}$}, the existence of the
non-decoupling top-quark corrections. It is a very important test of
the electroweak theory as a renormalizable quantum field theory with
spontaneous symmetry breakdown.

Finally, let us look into the bosonic contribution. It was pointed
out in \cite{DSKK} by using various high-energy data that such
bosonic electroweak corrections are now inevitable. I studied whether
we could observe a similar evidence in the $M_W$-$M_Z$ relation. In
this case, we have to compute $M_W$ taking account of only the
pure-fermionic corrections ${\mit\Delta}r[f](\equiv{\mit\Delta}r-
{\mit\Delta}r[{\rm boson}])$. Since ${\mit\Delta}r[f]$ depends on
$m_t$ strongly, it is not easy to develop a quantitative analysis of
it without knowing $m_t$. Therefore, I used the CDF data on $m_t$. I
express thus-computed $W$-mass as $M_W[{\rm f}]$. The result became
\begin{eqnarray}% ---------------------------------------------------
M_W[{\rm f}]=80.44\pm 0.11\ {\rm GeV}.
\end{eqnarray}% -----------------------------------------------------
This value is of course independent of the Higgs mass, and leads to
\begin{eqnarray}% ---------------------------------------------------
M_W[{\rm f}]-M_W^{exp}=0.21\pm 0.21\ {\rm GeV}, \label{eq2}
\end{eqnarray}% -----------------------------------------------------
which tells us that some non-fermionic contribution is necessary at
$1\sigma$ level.

It is of course too early to say from Eq.(\ref{eq2}) that the bosonic
effects were confirmed. Nevertheless, this is an interesting result
since we could observe nothing before: Actually, the best information
on $m_t$ before the CDF report was the bound $m_t^{exp} >$ 131
GeV by D0 \cite{D0}, but we can thereby get only $M_W[{\rm f}] >$
80.19 ($\pm$0.03) GeV (i.e., $M_W[{\rm f}]-M_W^{exp} > -0.04\pm 0.18$
GeV). We will be allowed therefore to conclude that ``the bosonic
effects are starting to appear in the $M_W$-$M_Z$ relation''.

We have seen that the standard electroweak theory seems now very
happy. Isn't there any problem in this theory, then? Najima and I
pointed out one thing in \cite{ZH942}. I showed that the $W$-mass
with the whole corrections for $m_t^{exp}=174\pm 17$ GeV and
$m_{\phi}=300$ GeV is consistent with the data. However, in order for
$M_W|_{m_t=174\ {\rm GeV}}$ to reproduce the central value of
$M_W^{exp}$ (80.23 GeV), the Higgs mass needs to be 1.1-1.2 TeV
\cite{ZH942}. Even if we limit discussions to perturbation
calculations, such an extremely-heavy Higgs will cause several
problems \cite{DM,DKR}. Moreover, the present LEP and SLC data
require a light Higgs boson: $m_{\phi}$
$\lower0.5ex\hbox{$\buildrel <\over\sim$}$ 300 GeV \cite{EFL}. This
means that we might be caught in a kind of dilemma.

At present, it is never serious since $m_{\phi}$ as low as 60 GeV is
also allowed if we take into account
${\mit\Delta}m_t^{exp}=\pm 17$ GeV and ${\mit\Delta}M_W^{exp}
=\pm 0.18$ GeV ($M_W-M_W^{exp}=0.20\pm 0.21$ GeV for $m_{\phi}=60$
GeV). Still, this definitely shows that more precise measurements
of $M_W$ and $m_t$ are considerably significant not only for
precision tests of the electroweak theory but also for new-physics
searches beyond this theory.
%--------------------------------------------------------------------
%\newpage

\vspace*{0.3cm}
\centerline{ACKNOWLEDGEMENTS}

\vspace*{0.3cm}
I am grateful to R. Najima for collaboration in \cite{ZH942}, on
which a part of this talk is based. I also would like to thank
S. Matsumoto for stimulating discussions on the data of $M_W$.

%\vskip 0.8cm

\end{document}